# The enzyme-like catalytic activity of cerium oxide nanoparticles and its dependency on $Ce^{3+}$ surface area concentration

V. Baldim*[1], F. Bedioui[2], N. Mignet[2], I. Margaill[3] and J.-F. Berret*[1]

[1]*Matière et Systèmes Complexes, UMR 7057 CNRS Université Denis Diderot Paris-VII, Bâtiment Condorcet, 10 rue Alice Domon et Léonie Duquet, 75205 Paris, France*
[2]*Chimie ParisTech - PSL Research University, INSERM U1022, CNRS 8258, Université Paris Descartes - Sorbonne Paris Cité, Unité de Technologies Chimiques et Biologiques pour la Santé, (UTCBS), 75006 Paris, France*
[3]*EA 4475, Pharmacologie de la Circulation Cérébrale, Université Paris Descartes, Faculté de Pharmacie de Paris, 4 avenue de l'Observatoire, 75006 Paris, France*

**Abstract:** Cerium oxide nanoparticles are known to catalyze the decomposition of reactive oxygen species such as superoxide radical and hydrogen peroxide. Herein, we examine the superoxide dismutase (SOD) and catalase (CAT) mimetic catalytic activities of nanoceria and demonstrate the existence of generic behaviors. For particles of size 4.5, 7.8, 23 and 28 nm, the SOD and CAT catalytic activities exhibit the characteristic shape of a Langmuir isotherm as a function of cerium concentration. Results show that the catalytic effects are enhanced for smaller particles and for the particles with the largest $Ce^{3+}$ fraction. The SOD-like activity obtained from the different samples is found to superimpose on a single master curve using the $Ce^{3+}$ surface area concentration as a new variable, indicating the existence of particle independent redox mechanisms. For the CAT assays the adsorption of $H_2O_2$ molecules at the particle surface modulates the efficacy of the decomposition process and must be taken into account. We design an amperometry-based experiment to evaluate the $H_2O_2$ adsorption at nanoceria surfaces, leading to the renormalization of the particle specific area. Depending on the particle type the amount of adsorbed $H_2O_2$ molecules varies from 2 to 20 $nm^{-2}$. The proposed scalings are predictive and allow determining SOD and CAT catalytic properties of cerium oxide solely from physico-chemical features.

Corresponding authors:
victor.baldim@univ-paris-diderot.fr, jean-francois.berret@univ-paris-diderot.fr

# Introduction

Cerium oxide nanoparticles (nanoceria) exhibit a fluorite-like crystal structure with a face-centered-cubic unit cell and are made of divalent oxygen anions $O^{2-}$ and tetravalent cerium cations $Ce^{4+}$. Defects in the crystal lattice are associated with oxygen vacancies surrounded by two cerium ions in their reduced trivalent state, $Ce^{3+}$. The coexistence of two oxidation states confers to these particles remarkable antioxidant and catalytic properties.[1-3] Nanoceria are hence non-stoichiometric particles and designated by $CeO_{2-x}$, where $2x$ is the $Ce^{3+}$ fraction.

Nowadays, cerium oxide nanoparticles (CNPs) are present in several industrial processes and products. Applications include chemical-mechanical polishing, UV filtering and fuel cells, but also heterogeneous catalysis for the treatment of environmental contaminants and the reduction of diesel particulate emissions. In general, the catalytic processes consist in the adsorption of reactants at the nanoceria surface, followed by surface reaction of the adsorbed species and by the desorption of products.[4,5] More recently, researchers have suggested that nanoceria catalytic activity may find applications in biology, primarily to scavenge exogenous reactive oxygen species (ROS), as recently reviewed by Xu and Qu.[6] The targeted ROS molecules are superoxide radical anions ($O_2^{-\bullet}$), hydrogen peroxide ($H_2O_2$), hydroxyl radicals ($OH^\bullet$), nitric oxide (NO) and peroxynitrite ($ONOO^-$).[7-10] In this context, nanoceria dispersions were assessed *in vitro* over a large variety of cell lines, e.g. lung epithelial cells, retinal neurons, colon cells, macrophages and cancer cell lines.[11-18] In most cases, nanoceria were found to act as an antioxidant by controlling the ROS level in cultures and by diminishing the cell mortality. In fewer cases, relevant pro-oxidant results were also reported.[2,6,19] Some authors studied CNPs *in vivo* in relation to pathologies associated with oxidative stress and ROS production. It was shown for instance that nanoceria can protect photoreceptor cells present in the retina and avoid peroxide-induced retinal tissue degeneration.[20] Nanoceria beneficial effects were also found in cardiomyopathy treatment,[21] in the regulation of the inflammatory responses in sepsis[22] and in the relief of multiple sclerosis progression.[23] More recently, Kim *et al*. have brought evidences that ceria nanoparticles can be also neuroprotective and diminish the infarction volume in cerebral ischemic stroke.[24] These findings strongly support the idea that CNP properties, both *in vitro* and *in vivo* are due to their ability to protect healthy cells and tissues against deleterious ROS effects. However, the mechanisms behind this catalytic activity are not yet fully understood and remain a matter of debate.

According to the early works of Baer and Stein,[25] Siegler and Masters[26] and Bielski and Saito[27] in the late 1950's, the reduction of $Ce^{4+}$ aqueous solutions by $H_2O_2$ takes place in a two steps redox mechanism: $Ce^{4+}$ is first reduced by $H_2O_2$ into $Ce^{3+}$ while hydroperoxyl radicals ($HO_2^\bullet$) are produced. In a second reaction these radicals are also involved in the reduction of $Ce^{4+}$ into $Ce^{3+}$. The overall reaction products are then $Ce^{3+}$ ions, protons, water and molecular oxygen.[25-28] In their pioneering work, Kitajima *et al*. reported that $H_2O_2$ does react with supported metal oxide substrates[29] and produce the superoxide species $O_2^{-\bullet}$. Subsequently, Seal and coworkers proposed that nanoceria dismutate $O_2^{-\bullet}$ and disproportionate $H_2O_2$ in a similar way as the enzymes superoxide dismutase (SOD) and catalase (CAT). These authors suggested that the SOD-like activity of nanoceria was mainly dependent on the $Ce^{3+}$ fraction, while the CAT-like activity would depend on the $Ce^{4+}$ fraction.[7,8,30] By means of high-energy resolution X-ray spectroscopy, Cafun *et al*. proposed the "electron sponge" model in which delocalized negative charges over the entire particle volume is relevant to account for nanoceria chemical activity.[31] Several groups also examined the effect of the size, morphology and surface coating on the reactivity of CNPs towards $H_2O_2$ and found that smaller particles have higher $Ce^{3+}$ fractions in general and are also more reactive.[32-34] Finally, recent studies supported the idea that oxidation of $Ce^{3+}$ into $Ce^{4+}$ induced by $H_2O_2$ leads to the formation of stable intermediate species, such as peroxo or hydroperoxo species bound to the nanoceria surface.[35-38]

The picture that emerges from the previous studies is that the mechanisms underlying nanoceria catalytic activities strongly depend on the particle properties (size, morphology, $Ce^{3+}$ fraction etc…), but a quantitative description of this dependence is still missing. In this work, we measured the SOD and CAT-like activities of four different particles of sizes 4.5, 7.8, 23 and 28 nm over extended cerium concentrations. These measurements reveal that for all particles the catalytic activities exhibit the characteristic shape of a Langmuir adsorption isotherm when plotted as a function of the Ce concentration. Moreover, the catalytic activity can be rescaled using reduced variables, leading to a generic behavior. For the SOD assays, the reduced variable involved in the scaling is identified as the $Ce^{3+}$ surface area concentration. For the CAT assays we found that the adsorption of $H_2O_2$ molecules at the particle surface modulates the efficacy of the disproportionation process and should be taken into account in the scaling. To achieve this goal, we design an amperometry experiment and evaluate the adsorption of hydrogen peroxide at nanoceria surfaces. These assays provide the adsorption kinetics as well as the $H_2O_2$ densities at the surface prior to its disproportionation.

## II - Results and discussion
### II.1 - Particle characterization

Transmission electron microscopy (TEM), X-ray photoelectron spectrometry (XPS) and electronic ultraviolet-visible spectroscopy (UV-vis) are used to characterize the nanoceria samples investigated in this work. Figs. 1a-d show representative TEM micrographs of the four nanoceria samples, together with the size distributions (insets). The size distributions are adjusted using log-normal functions, leading to median diameters of 4.5, 7.8, 23 and 28 nm respectively. In the following, the cerium oxide dispersions will be named CNP-5, CNP-8, CNP-23 and CNP-28 in reference to their TEM sizes. The images in Fig. 1 and those in Supplementary Information (Fig. S1) illustrate that CNP-5 particles are spherical nano-objects, whereas CNP-8 consist of small agglomerates of 2 nm crystallites. These latter findings are in agreement with our earlier determinations on this sample.[39-41] For the CNP-23 and CNP-28 samples, the particles are polyhedral and characterized by a broad size dispersity. The specific surface area $A_S$ of CNP-5, CNP-8, CNP-23 and CNP-28 nanoparticles is calculated from their median diameters and dispersities, according to:

$$A_S = \frac{6\exp(-2.5s^2)}{D\rho} \quad (1)$$

where $s$ is the dispersity, $D$ the diameter and $\rho$ the mass density. Details for the calculations are provided in Supplementary Information S2. $A_S$ is found to be 136, 101, 24 and 24 $m^2\,g^{-1}$, respectively (Table I).

Studies have shown that besides the size, an important feature affecting nanoceria antioxidant properties is its $Ce^{3+}$-fraction, noted $f_{Ce^{3+}}$ in the following.[42] To determine $f_{Ce^{3+}}$, XPS experiments are performed on powder samples obtained after drying the particle dispersions. Fig. 1e shows the decomposed Ce3d XPS spectra for the four nanoparticles. Three pairs of peaks are associated to Ce3d electrons ejected from $Ce^{4+}$, while two others pairs of peaks are due to $Ce^{3+}$ ions. The complete analysis of the XPS spectra including the peak assignment, the determination of the binding energies and of the $Ce^{3+}$ fraction is provided in the Supplementary Information (S3). For the CNP-5, CNP-8 CNP-23 and CNP-28 one gets $f_{Ce^{3+}}$ = 43%, 14%, 9.0% and 9.5% respectively. This fraction is found to increase with decreasing nanoparticle size, a result that was noticed in earlier studies and that is related to the lattice strain induced by $Ce^{3+}$ and the presence of oxygen vacancies.[43] Wide-angle X-ray scattering (WAXS) performed on the 4 samples confirms the fluorite-like structure of the nanocrystals (Supplementary Information S4). From the Rietveld MAUD analysis, the lattice constants were derived and found to increase linearly with the $Ce^{3+}$ fraction, a result that is attributed to the decrease of electrostatic force caused by valence reduction from $Ce^{4+}$ to $Ce^{3+}$ ions.[43] The overall characteristics of the cerium oxide samples are summarized on Table I.

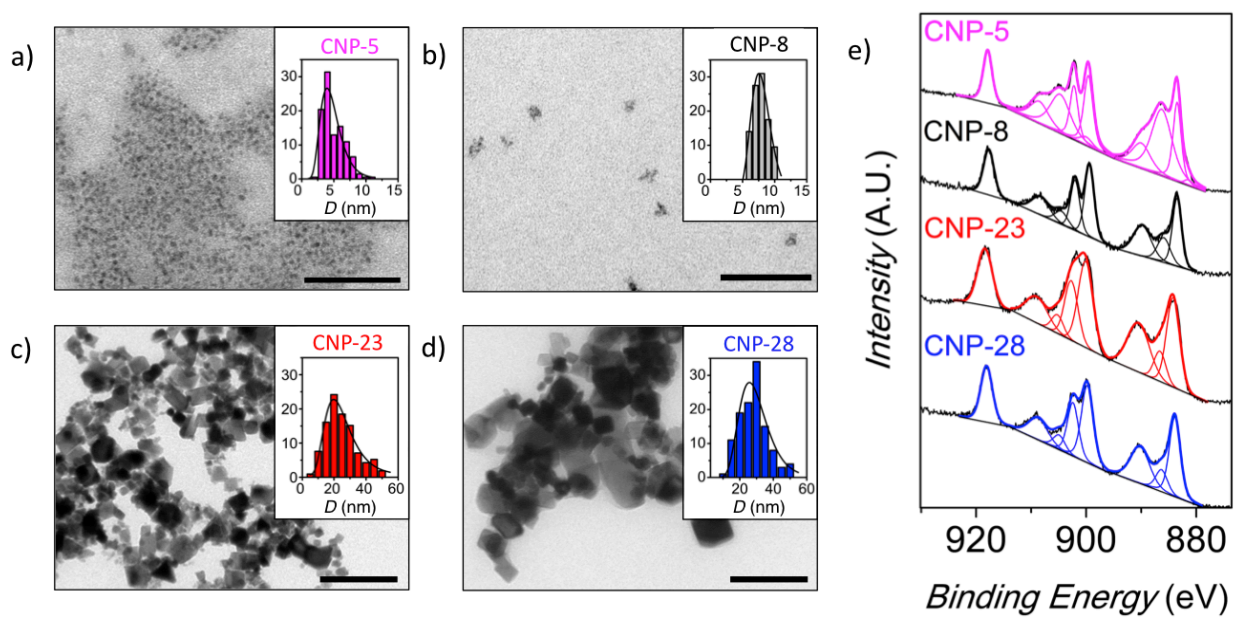

*Figure 1: Electron microscopy and X-ray photoelectron spectrometry results*

*Nanoceria characterization using transmission electron microscopy (a-d) and X-ray photoelectron spectrometry (e). For the TEM micrographs, the scale bars are 50 nm for (a) and (b) and 100 nm for (c) and (d). Insets: size distributions obtained from the analysis on n = 200 particles. The distributions are adjusted using a log-normal function with median diameter D and dispersity s (see Supplementary Information S2 for details). The dispersity is defined as the ratio between the standard deviation and average size.[44] e) XPS Ce3d spectra for CNP-5 (magenta), CNP-8 (black), CNP-23 (red) and CNP-28 (blue). The continuous thick lines display the sum of the different peak contributions adjusted according to the model described in S3.*

| $CeO_{2-x}$ nano-particles | $D$ nm | $s$ | $a$ nm | Morphology | $A_S$ $m^2 g^{-1}$ | $f_{Ce^{3+}}$ % |
|---|---|---|---|---|---|---|
| CNP-5 | 4.5 | 0.36 | 0.544572 | Spherical | 136 | 43 |
| CNP-8 | 7.8 | 0.17 | 0.541514 | Agglomerate | 101 | 14 |
| CNP-23 | 23 | 0.42 | 0.540858 | Polyhedral | 24 | 9.0 |
| CNP-28 | 28 | 0.31 | 0.540999 | Polyhedral | 24 | 9.5 |

*Table I - Nanoceria characteristics.*
*The median diameter (D), the dispersity (s) and the morphology were obtained from transmission electron microscopy measurements. The lattice constant (a) of the cerium oxide nanocrystals and the $Ce^{3+}$ fractions ($f_{Ce^{3+}}$) were determined from wide-angle X-ray scattering and X-ray photoelectron spectrometry respectively. The dispersity is defined as the ratio between the standard deviation and average size.*

Another remarkable property displayed by nanoceria is its propensity to absorb UV light and to quickly change color when mixed with hydrogen peroxide ($H_2O_2$).[42,45] Fig. 2 shows UV-vis absorption spectra and photographs of CNP-5, CNP-8, CNP-23 and CNP-28 dispersions before and after $H_2O_2$ addition. The spectral and visual color changes are much more intense for the smaller nanoparticles than for the larger ones, a result that was already noticed by many authors.[33] Following Lee *et al.*,[42] we characterize the observed red shift by measuring the wavelength difference $\Delta\lambda$ at the optical density 0.3 before and after peroxide addition and found $\Delta\lambda$ = 90, 100, 17 and 10 nm for CNP-5, CNP-8 CNP-23 and CNP-28 respectively. This shift is regarded as a measure of the NP antioxidant capacity and allows simple comparison between samples. The colorimetric data show that increasing the particle size decreases the shift of the spectrum, a result that was attributed to the oxidation of $Ce^{3+}$ surface ions into $Ce^{4+}$ by $H_2O_2$[42] or/and to coordinated peroxide species onto CNP surfaces.[35,37,46] Supplementary Information S5a displays absorbance results as a function of $H_2O_2$ (from 0 to 882 mM) and confirms the saturation of the red shift at high concentrations.[34,37,42] Additional colorimetric assays on the absorbance kinetics after addition of hydrogen peroxide bring evidences of nanoceria catalytic cycling properties after a 6 day incubation (S5b). The results of the previous sections demonstrate that the nanoceria studied here display similar structural, chemical and optical properties than those of the literature. They were selected to cover a wide range of physico-chemical features. In the next section we focus on their superoxide dismutase and catalase mimetic activities.

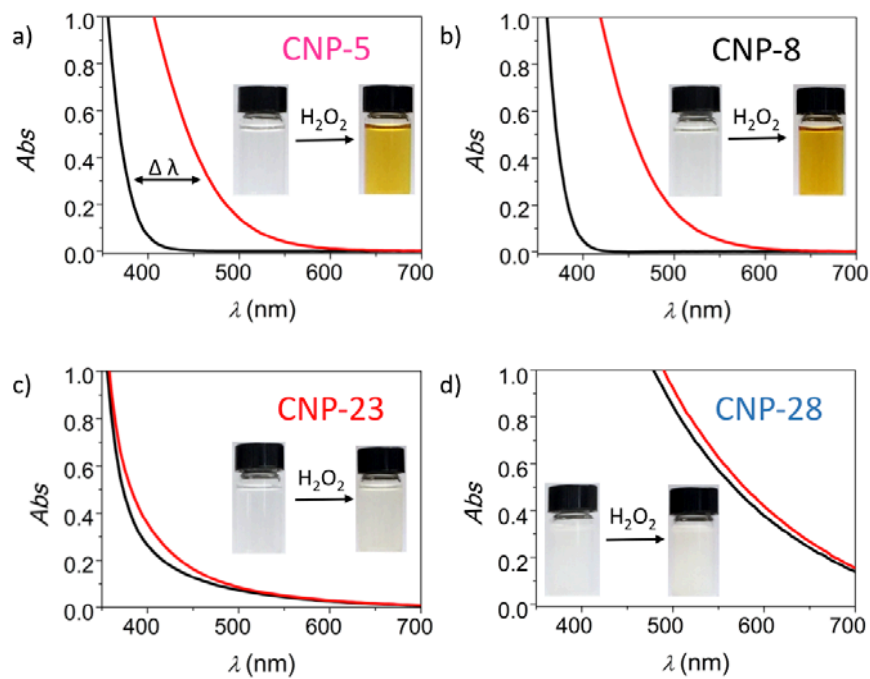

*Figure 2: Absorbance properties of nanoceria with and without hydrogen peroxide*
*UV-vis absorption spectra of nanoceria dispersions at the concentration of 1.6 mM before (black lines) and two minutes after addition of a concentrated hydrogen peroxide to the final concentration of 60 mM (red lines). a) CNP-5; b) CNP8; c) CNP-23 and; d) CNP-28. Insets: images of the nanoceria dispersions with and without hydrogen peroxide. Only the smallest particles exhibit significant changes in the range 300 – 500 nm.*

### II. 2 - Superoxide dismutase mimicking activity of nanoceria
The SOD-like catalytic activity of nanoceria was investigated by a colorimetric assay using UV-Vis spectroscopy.[47,48] With this assay, superoxide radical anions are generated *in situ* by xanthine oxidase in a nanoceria dispersion. A water-soluble tetrazolium salt is then added to the solution, this latter being oxidized by the remaining superoxide radicals into a soluble formazan dye. The final product is then identified from its absorption property.[24,47,49] The SOD-like activity $A_{SOD}$ is defined as the percentage of dismutated superoxide radicals at the end of the assay. Fig. 3a shows the effect of the particle type on $A_{SOD}$ at cerium concentration 200 µM. $A_{SOD}$ is found to decrease with increasing particle diameters, suggesting a potential size and $Ce^{3+}$ fraction effect. Fig. 3b illustrates the concentration dependence of $A_{SOD}$ in the range 200 - 2000 µM. Increasing the CNP concentration leads to increased dismutation rates for the 4 samples. To account for this dependence, the SOD results are displayed *versus* the surface area concentration (expressed in m$^2$ L$^{-1}$):

$$c_S = A_S c \qquad (2)$$

where the specific surface area $A_S$ (Eq. 1) is in m$^2$ g$^{-1}$ and the weight concentration in g L$^{-1}$ (see also Supplementary Information S2). Fig. 3c shows that the SOD activity scales well for CNP-8, CNP-23 and CNP-28, the data being now well superimposed. This superposition is however not observed for CNP-5, *i.e.* for the particles with the highest $f_{Ce^{3+}}$. These data suggest that the surface density of $Ce^{3+}$ active sites plays a significant role on the SOD response.[7] Fig. 3d retraces the previous data as a function of the surface area concentration *times* the $Ce^{3+}$ fraction ($c_S f_{Ce^{3+}}$). It is found that the SOD-like activity of the different nanoceria samples are well superimposed and follows a sigmoid-type behavior when plotted in semi-logarithmic scale (continuous curve in grey in Fig. 3d). These data are adjusted by a Langmuir-type adsorption isotherm of the form:[50]

$$A_{SOD}(c_S, f_{Ce^{3+}}) = \frac{1}{1 + c_0^{SOD}/c_S f_{Ce^{3+}}} \qquad (3)$$

where $c_0^{SOD}$ (= 5.6 m$^2$ L$^{-1}$) is the only adjustable parameter. Fig. S6 illustrates the Langmuir isotherm in linear scale together with the Eq. 1, showing in particular the existence of a linear regime at low $c_S f_{Ce^{3+}}$ followed by a saturation plateau. The findings of Fig. 3 substantiate the existence of a superposition principle once the surface area concentration and the $Ce^{3+}$ fraction are taken into account.[7,42,43] Here the multivariable dependencies of the SOD mimetic activity have been reduced into a single master curve and the $Ce^{3+}$ surface area concentration $c_S f_{Ce^{3+}}$ has been identified as the key parameter for the superposition principle. We suggest that this result is quite general and could be used to predict the catalytic activity of nanoceria, knowing its particle size and $Ce^{3+}$ fraction.

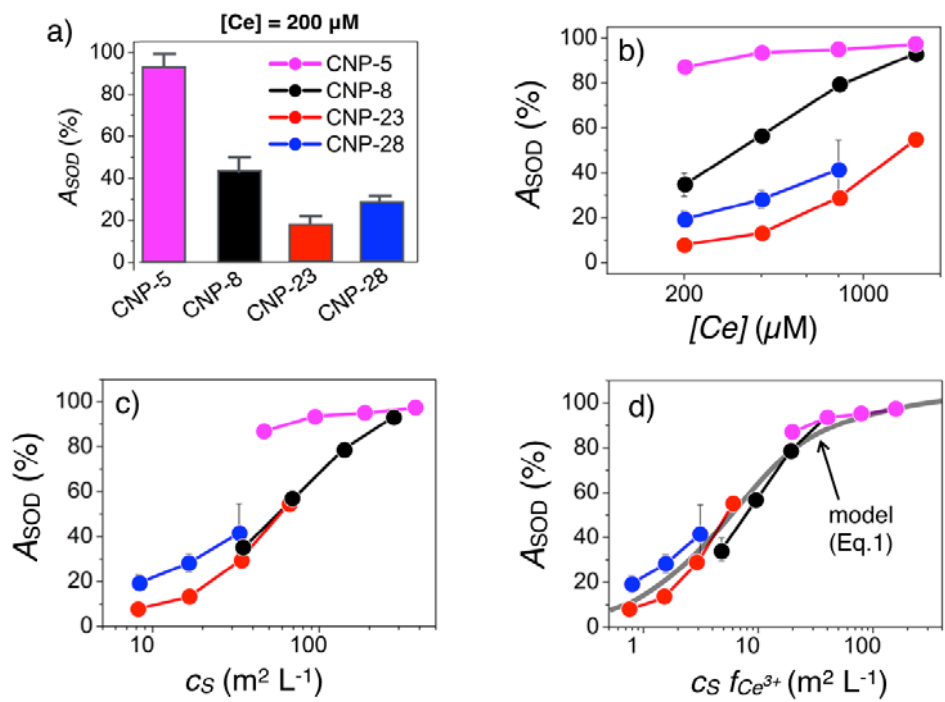

*Figure 3: Superoxide dismutase-like activity of nanoceria*
*a) Percentages of dismutated superoxide radicals $A_{SOD}$ obtained for CNP-5 (magenta), CNP-8 (black), CNP-23 (red) and CNP-28 (blue) at cerium concentration 200 µM. In figures b, c and d, $A_{SOD}$ is plotted respectively as a function of the cerium molar concentration [Ce], of the surface area concentration $c_S$ and of the surface area concentration times $Ce^{3+}$ fraction, $c_S f_{Ce^{3+}}$. In d), the continuous line is obtained from Eq. 1 using $c_0^{SOD} = 5.6$ $m^2$ $L^{-1}$ as a single adjustable parameter (see also S7).*

### II. 3 - Catalase-like catalytic activity

The nanoceria catalase-like catalytic activity was investigated by spectrofluorimetry using the Amplex-Red reagent assay.[22,24,47] In this assay, cerium oxide dispersions at molar concentration $3\times10^{-1}$ to $3\times10^{4}$ µM are incubated with 5 µM hydrogen peroxide in 96-well plates. A mixture of horseradish peroxidase and Amplex-Red is then added, this latter being transformed into the fluorescent resorufin from the remaining $H_2O_2$.[24] The CAT-like activity $A_{CAT}$ is defined as the percentage of decomposed $H_2O_2$ at the end of the assay. Fig. 4a shows the effect of the particle size on $A_{CAT}$ at cerium concentration 200 µM. $A_{CAT}$ is found to be smaller for the large particles CNP-23 and CNP-28, suggesting again a potential size effect. When plotted in semi-logarithmic scale against the added cerium oxide, $A_{CAT}$ exhibits a sigmoidal-shape dependence similar to that found for SOD (Fig. 4b). We also note some differences between the CNPs. For the smallest particles (CNP-5 and CNP-8), the $H_2O_2$ percentages increase at concentration around 1 µM and display comparable responses. For CNP-23 and CNP-28, the sigmoids are shifted to higher concentrations, with a signal onset starting around 100 µM and a saturation above $10^4$ µM. Following the approach developed in the previous section, the $H_2O_2$ disproportionation results have been rescaled using first the specific surface concentration $c_S$ (Fig. 4c) and second using the specific surface concentration *times* the $Ce^{3+}$ fraction, $c_S f_{Ce^{3+}}$ (Fig. 4d). After rescaling, the set of $A_{CAT}$-curves obtained have neared closer to each other, indicating that the size and the $Ce^{3+}$ fraction are here too key parameters in the activity of cerium oxide. The sigmoidal behavior was fitted by a Langmuir-type adsorption function of the form:[50]

$$A_{CAT}(c_S, f_{Ce^{3+}}) = \frac{1}{1 + c_0^{CAT}/c_S f_{Ce^{3+}}} \qquad (4)$$

where $c_0^{CAT}$ is the only adjustable parameter. In Supplementary Information S7 we show that the agreement between the results and Eq. 2 is excellent. The $c_0^{CAT}$ values retrieved from the fitting are 0.57, 0.11, 0.61 and 3.57 m$^2$ L$^{-1}$ for CNP-5, CNP8, CNP-23 and CNP-28 respectively. In contrast to the SOD-like activity, the data here do not superimpose, suggesting the existence of additional contributing factors.

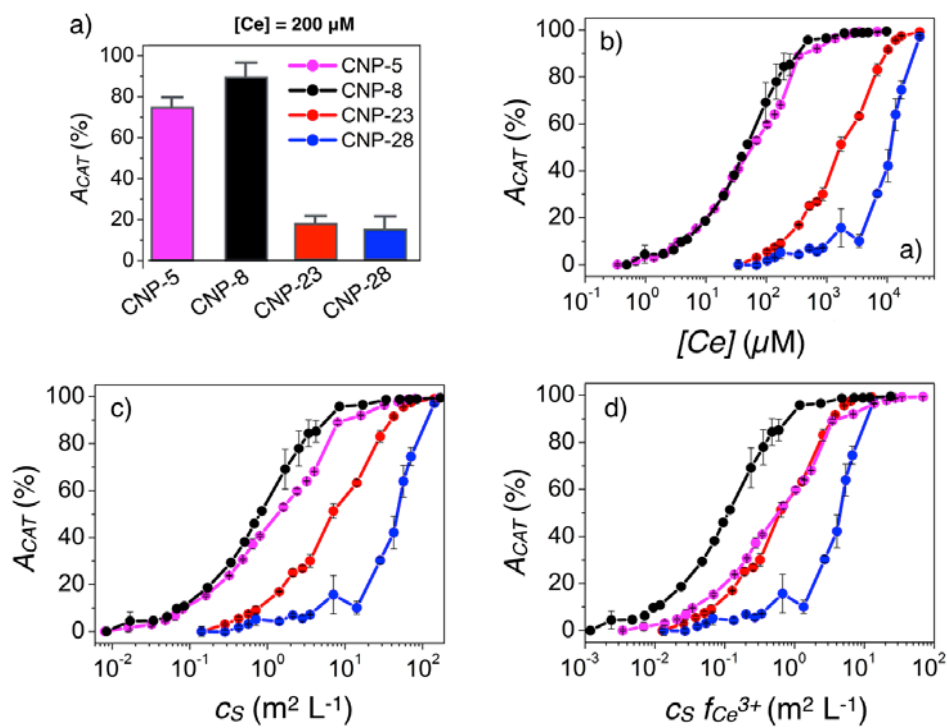

*Figure 4: Catalase-like catalytic activity of nanoceria*
*a) Percentages of disproportionated hydrogen peroxide $A_{CAT}$ obtained for CNP-5 (magenta), CNP-8 (black), CNP-23 (red) and CNP-28 (blue) at cerium concentration 200 µM. In figures b, c and d, $A_{CAT}$ is plotted respectively as a function of the cerium molar concentration [Ce], of the surface area concentration $c_S$ and of the surface area concentration times Ce$^{3+}$ fraction, $c_S f_{Ce^{3+}}$. In these assays, the initial H$_2$O$_2$ concentration was 5 µM.*

## II. 4 - Hydrogen peroxide adsorption at the nanoceria surface

To get some insight into the process occurring upon mixing H$_2$O$_2$ and nanoceria,[35,37,38] we designed an amperometry experiments capable of monitoring the H$_2$O$_2$ concentration based on the changes in electric currents. Briefly, a working Pt/Pt-black ultramicroelectrode (UME) and a reference Ag/AgCl electrode are placed inside a microplate well containing Tris-Cl buffer pH 7.5 and connected to a potentiostat.[51,52] A potential of 0.4 V is applied and the current is recorded while hydrogen peroxide and nanoceria dispersions are added consecutively to the solution. The representation in Fig. 5a illustrates the experimental set-up and provides the working conditions for the two electrodes. Fig. 5b shows the time evolution of the current during a typical amperometry monitoring. A base line is first recorded and is followed by a step increase of the current at 1 nA after $H_2O_2$ addition. Five minutes later, CNPs are added to the previous solution, leading to a rapid decrease of the current and to its stabilization at long time. Fig. 5c displays further measurements aiming at calibrating the UME. Different volumes of a 15 mM $H_2O_2$ solution are added stepwise to the initial solution, leading to steps in the current response. Similarly, additions of nanoceria (CNP-8, 60 mM) produce a step-like diminution of the electric signal in the proportion of the added volume. From the first part of the assay, a linear current-concentration calibration characterized by a slope of 0.011 nA µM$^{-1}$ is observed (Fig. 5d). The rapid decrease found upon $CeO_{2-x}$ addition is caused by the decrease of the $H_2O_2$ concentration consecutive to the adsorption of hydrogen peroxide at the nanoceria surface (Fig. 5e). It is followed by the formation of adsorbed oxygen species such as peroxide or hydroperoxide.[35-38] To our knowledge, this is the first report where chrono-amperometry was exploited to investigate and quantify such properties. Liu *et al*. also reported the adsorption of hydrogen peroxide to nanoceria surfaces and exploited it to develop a DNA fluorescence-based method for hydrogen peroxide titration.[53]

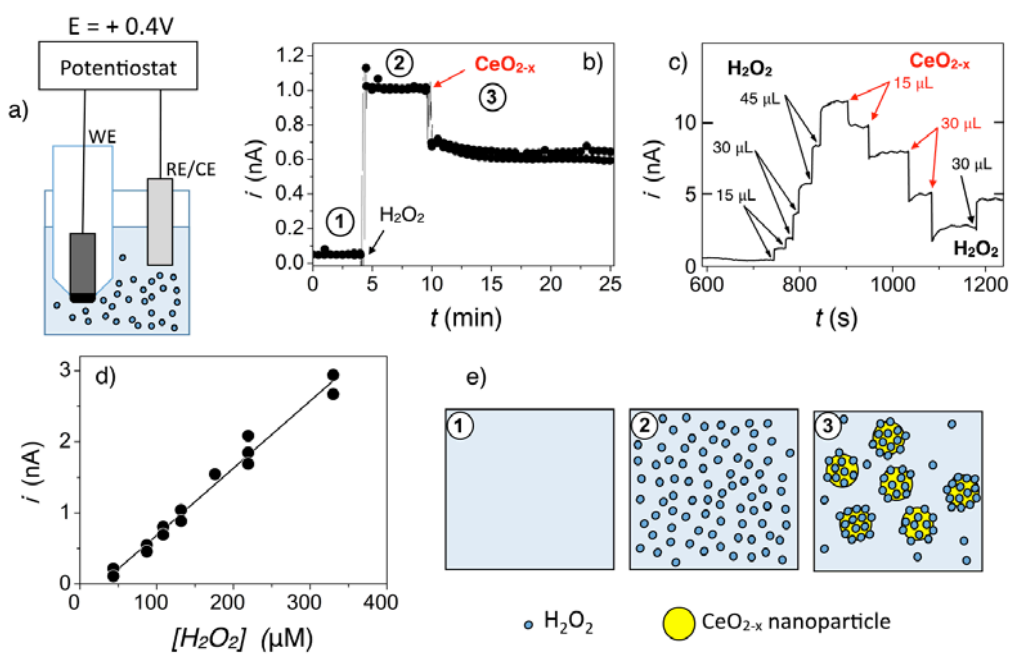

*Figure 5: Amperometric detection of hydrogen peroxide in the presence of nanoceria*
*a) Experimental set-up used for the amperometric monitoring of hydrogen peroxide including the Pt/Pt-black working ultramicroelectrode and the Ag/AgCl reference electrode. b) Typical time evolution of the current upon $H_2O_2$ and $CeO_{2-x}$ addition. Note that the characteristic time scale for the [$H_2O_2$] decrease is similar to that of the absorbance changes seen in UV-Vis spectrometry. c) Signal measurements following the addition of increasing volumes of a 15 mM $H_2O_2$ solution and of a 60 mM CNP-8 dispersion, as indicated. d) Calibration curve between hydrogen peroxide concentration and current. The straight line is the result of least square adjustment, providing a slope of 0.011 nA µM$^{-1}$. e) Schematic representation of hydrogen peroxide adsorption at nanoceria surfaces.*

To compare the CNPs with the amperometry technique, dispersions were prepared at fixed surface area concentration $c_S$ = 10 m$^2$ L$^{-1}$. Upon CNP addition, all experiments reveal a decrease of the $H_2O_2$ concentration, as previously mentioned (Fig. 6a). This decrease depends however on the sample type, suggesting different adsorption site densities. The signal reduction is for instance stronger for CNP-8 compared to that of CNP-5. We estimate the surface density of hydrogen peroxide molecules from these experiments and find 12.0 ± 0.7 nm$^{-2}$, 16.7 ± 0.8 nm$^{-2}$, 5.3 ± 0.1 nm$^{-2}$ and 2.3 ± 0.5 nm$^{-2}$ for CNP-5, CNP-8, CNP-23 and CNP-28 respectively. These values are now used to define an effective specific surface for the CAT assay, noted $A_{ES}$. Here, we consider CNP-8 as a reference and recalculate the effective surface area density of active sites for the other particles using the $A_{ES}$ values given in Table II. Fig. 6b displays the $A_{CAT}$ percentages as a function of the effective surface concentrations $c_{ES}$ *times* the $Ce^{3+}$ fraction in semi-logarithmic scale. The graph shows a good superimposition of the different curves, indicating that the nanoceria CAT-like activity depends mainly on three factors, the particle size, the $Ce^{3+}$ fraction and the $H_2O_2$ adsorption properties. Replacing the concentration $c_S$ by the effective concentration $c_{ES}$, the data in Fig. 6 are adjusted using Eq. 2 and the parameters $c_0^{CAT}$ are found to be 0.41, 0.11, 0.20 and 0.39 m$^2$ L$^{-1}$ with increasing particle sizes. The persisting differences between samples can be attributed to the approximations made estimating the scaling factors and in particular the specific surface area. $A_S$ was indeed calculated assuming that the particles are spherical objects. Another factor could be the dependence of the catalytic activity on the exposed crystalline facets.[3,34,54-56] The proposed superposition principle found here could be used to determine the CAT mimetic properties from the knowledge of the parameters mentioned above.

| $CeO_{2-x}$ nano-particles | Δ[$H_2O_2$] μM | Adsorbed $H_2O_2$ molecules nm$^{-2}$ | Effective surface $A_{ES}$ m$^2$ g$^{-1}$ |
|---|---|---|---|
| CNP-8 | 262 ± 13 | 16.7 ± 0.8 | 101 |
| CNP-5 | 189 ± 11 | 12.0 ± 0.7 | 98 |
| CNP-23 | 85 ± 2 | 5.3 ± 0.1 | 8 |
| CNP-28 | 35 ± 8 | 2.3 ± 0.5 | 3 |

**Table II**: Relevant parameters of the amperometric monitoring of hydrogen peroxide in the presence of nanoceria.

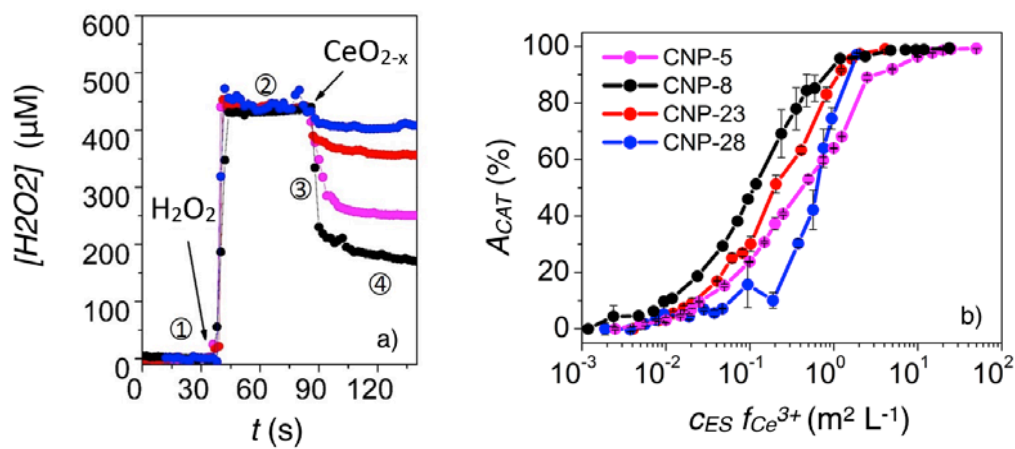

*Figure 6: Catalase-like activity of nanoceria in rescaled units*
*a) Chrono-amperogram showing the evolution of hydrogen peroxide concentration after and before addition of hydrogen peroxide (440 μM) followed by addition of CNP-5, CNP-8, CNP-23 or CNP-28*

*dispersions at the same surface per volume concentration (10 $m^2$ $L^{-1}$). b) Percentages of disproportionated $H_2O_2$ as a function of the effective surface concentrations times the $Ce^{3+}$ fraction in semi-logarithmic scale. The data adjusted using Eq. 2 and the parameters $c_0^{CAT}$ is found to be 0.4, 0.2, 0.1 and 0.5 $m^2$ $L^{-1}$ for CNP-5, CNP-8, CNP-23 and CNP-28 respectively (S7).*

## III - Conclusion

In this work we examined the enzyme mimicking activity of a series of cerium oxide nanoparticles of different specific surface areas (from 20 to 140 $m^2$ $g^{-1}$) and of different $Ce^{3+}$ fractions (from 9 to 43%). The physico-chemical features of the CNPs were obtained by extensive TEM, XPS, WAXS and UV-Vis measurements and analyze quantitatively. The superoxide dismutase and catalase-like activities were investigated *versus* cerium oxide concentration *via* the xanthine oxidase and Amplex-Red reagent assays respectively. Results show that the catalytic effects are enhanced for smaller particles and for the particles with the largest $Ce^{3+}$ fraction. In an effort to identify the key factors affecting the nanoceria catalytic activity and to provide a rationale about these dependencies, various types of scaling analysis were tested. We have found that once expressed in terms of the $Ce^{3+}$ surface area concentration, the percentages of dismutated superoxide radicals determined for the different CNPs were found to superimpose on a single master curve. Moreover, the data follow a Langmuir-type adsorption isotherm behavior. In particular, we could demonstrate the existence of a linear regime at low surface area concentration followed by a saturation plateau, the transition between the two regimes occurring around 5 $m^2$ $L^{-1}$. The observed Langmuir behavior indicates that the catalytic assays are probing surface phenomena related to the adsorption/desorption of test molecules. For the CAT-like activity, we found again Langmuir type isotherm variations for the 4 CNPs. However, the scaling using the $Ce^{3+}$ surface area concentration in abscissa failed and could not be observed at first. To get more insight in the interaction of $H_2O_2$ with nanoceria surfaces, amperometry experiments were performed to monitor the $H_2O_2$ concentration as a function of the time and in different working conditions. This technique provides the $H_2O_2$ adsorption kinetics upon CNP addition, as well as the densities of the hydrogen peroxide molecules at the surface in the stationary regime. We found that the $H_2O_2$ adsorption depends on the CNP used and that the specific surface area first derived from the TEM experiments ought to be renormalized. The surface density of hydrogen peroxide molecules was comprised between 2.3 $nm^{-2}$ and 16.7 $nm^{-2}$. With these modifications, the CAT-like activity of cerium oxide was re-examined and the data coming from different CNPs could be superimposed into a single master curve. In conclusion, we show that the SOD and CAT mimicking catalytic activity obeys simple and generic behaviors for the different nanoparticles tested. We suggest that these scaling behaviors could be used to evaluate the catalytic performance of nanoceria solely from the nanoparticles physical and chemical properties.

## IV - Materials and Methods

### IV.1 - Materials

CNP-5 and CNP-23 are aqueous dispersions at concentrations 200 g $L^{-1}$ and 100 g $L^{-1}$ (both at pH4) purchased from Sigma-Aldrich. The particles are positively charged, associated with zeta potentials ζ = +8 mV and +40 mV respectively, and according to the provider do not carry additives at their surface. CNP-8 originates from a 200 g $L^{-1}$ low pH aqueous dispersion (pH 1.5) synthesized by Rhodia (Centre de Recherche d'Aubervilliers, Aubervilliers, France).[11,57,58] As shown in earlier reports, the CNP-8 are positively charged at acidic pH ($\zeta$ = + 40 mV) and aggregate above pH2.[40,41] The colloidal stability is ensured by electrostatic repulsion arising from positive surface charges, as discussed in Ref.[59]. In such a case, the surfaces are devoid of ligands or macromolecules. CNP-28 is a cerium oxide benchmark material (code name NM-212) synthesized for the OECD program "Testing a representative set of manufactured nanomaterials".[60] It was supplied as a fine powder by the Institute for Health and Consumer Protection (IHCP, Joint Research Centre of European Commission, Italy). In the report issued by the European Commission, the CNP-28 were extensively characterized with respect to their surface properties. The BET measurement gives a value of the specific surface area of $A_S$ = 27.2 m$^2$ g$^{-1}$, in good agreement with the TEM determination ($A_S$ = 24 m$^2$ g$^{-1}$, see Table I). The positive zeta potential ($\zeta$ + = 33 mV) indicates the presence of positive charges at the surface. XPS results indicate that carbon and oxygen were detected, most likely associated with surface contamination or adsorbed carbon containing species.[60] Hydrogen peroxide (H$_2$O$_2$, 30 vol. %), hydrogen hexachloroplatinate(IV) solution, lead(II) acetate trihydrate and the superoxide dismutase kit assay (Kit #19160-1KTF) were bought from Sigma-Aldrich (Lyon, France). The Amplex® Red catalase assay kit (Cat # A22180) was purchased from Thermo Scientific (Illkirch, France). T

### IV.2 – Ultraviolet-visible Spectroscopy (UV-vis)

A UV-visible spectrometer (SmartSpecPlus from BioRad) was used to measure the absorbance of nanoceria aqueous dispersions. The absorbance is related to the nanoparticle concentration by the Beer-Lambert expression $Abs(\lambda) = \epsilon(\lambda)\ell c$ where $\ell$ is the optical length, $c$ the nanoparticle concentration, $\epsilon(\lambda)$ the molar absorption coefficient. Absorbance data were used to determine the cerium oxide concentration for each batch.[39]

### IV.3 – Zeta potential

Laser Doppler velocimetry was used to carry out the electrokinetic measurements of electrophoretic mobility and zeta potential ($\zeta$) with the Zetasizer Nano ZS equipment (Malvern Instruments, Worcestershore, United Kingdom). Light scattering and zeta potential measurements were performed in triplicate at 25 °C after an equilibration time of 120 s.[61]

### IV.4 - Transmission Electron Microscopy (TEM)

Micrographs were taken with a Tecnai 12 TEM operating at 80 kV equipped with a 1K×1K Keen View camera. Nanoceria dispersions were deposited on ultrathin carbon type-A 400 mesh copper grids (Ted Pella, Inc.). Micrographs were analyzed using ImageJ software for 200 particles. The particle size distributions are adjusted using a log-normal function of the form (Fig. 1):

$$p(d, D, s) = \frac{1}{\sqrt{2\pi}\beta(s)d} exp\left(-\frac{ln^2(d/D)}{2\beta(s)^2}\right) \quad (5)$$

In the previous equation, $\beta(s)$ is related to the size dispersity $s$ through the relationship $\beta(s) = \sqrt{\ln(1 + s^2)}$. $s$ is defined as the ratio between the standard deviation and the average diameter.[44] For $\beta < 0.4$, one has $\beta \cong s$.[39,62]

### IV.5 - X-ray Photoelectron Spectroscopy (XPS)

XPS data was collected using an Omicron Argus X-ray photoelectron spectrometer using a monochromated AlK$_\alpha$ (1486.6 eV) radiation source with a 300 W electron beam power. The emission of photoelectrons from the sample was analyzed at a takeoff angle of 45° under ultra-high vacuum conditions ($10^{-8}$ Pa). The spectra were collected at pass energy of 100 eV for the survey scan and 20 eV for the high resolution scans. The XPS spectra were fitted using the software XPSPEAK41, applying a fixed Gaussian/ Lorentzian ratio for peaks of the same spectrum and constraining the full-width at half-maximum (FWHM) of each doublet to be equal. The fraction of $Ce^{3+}$ was determined through the ratio between the integrated intensities of the four peaks corresponding to $Ce^{3+}$ divided by the integrated intensities of all ten peaks, as detailed in Supplementary Information.

### IV.6 – Wide-Angle X-ray Scattering (WAXS)

WAXS was carried out using an Empyrean (PANALYTICAL) diffractometer equipped with a multichannel PIXcel 3D detector and a Cu Kα X-ray source (1.54187 Å). Samples were deposited on a monocrystalline Si substrate, with a spinner movement (rotation time 1 s). A 1/16° divergence slit, a 1/8° anti-scatter slit and a 10 mm mask were installed before the samples. Typically, each pattern was recorded in the $\theta - \theta$ Bragg-Brentano geometry in the 10° - 20° 2$\theta$ range (0.0263° for 600 s), where 2$\theta$ denotes the scattering angle.

### IV.7 - Amperometric monitoring of $H_2O_2$

A homemade Pt/Pt-black ultramicroelectrode (UME) was used as the hydrogen peroxide sensor. A Pt wire (diameter 25 μm) was put into a closed end glass capillary and then heated under vacuum until the glass was melted around the wire. Then, the capillary was filled with a conductive lacquer and a copper wire was finally inserted for connection with the potentiostat. The UME surface was polished on emery paper with diamond solution and Pt-black was deposited on its surface by the reduction of a hydrogen hexachloroplatinate(IV) solution containing Lead(II) acetate trihydrate by applying a current density $J$ of -8 mA cm$^{-2}$ and a charge density $Q$ of -240 mC cm$^{-2}$. The Pt/Pt-black UME was connected to a potentiostat (eDAQ Pty Ltd.) and immersed in 2 mL of Tris-Cl buffer pH 7.5 placed inside a well of a 24-well cell culture plate. The voltage was set at +0.4 V *versus* an Ag/AgCl reference counter electrode. After the baseline stabilization, 50 μL of a $H_2O_2$ solution were added to the well, followed by the addition of 50 μL of a nanoceria dispersion, while the current was recorded in real time.

### IV.8 - SOD mimetic activity assay

The catalytic activity of nanoceria in the dismutation of superoxide radical anion was assessed by a colorimetric assay using UV-Vis spectroscopy (Kit #19160-1KTF).[47,48] Briefly, 20 μL of a nanoceria dispersion in Tris-Cl buffer pH 7.5 was added to a well of a 96-well plate and mixed with 200 μL of WST-1 (2-(4-Iodophenyl)-3-(4-nitrophenyl)-5-(2,4-disulfophenyl)-2H-tetrazolium, monosodium salt). The reaction was initiated with the addition of 20 μL of xanthine oxidase solution, prepared by mixing 5 μL of the enzyme in 2.5 mL of a dilution buffer provided. After incubating plate at 37°C for 20 min, the absorbance at 450 nm was measured using a microplate reader (EnSpire Multimode Plate Reader, Perkin Elmer). Final nanoceria dispersion concentration ranged from 200 to 2000 μM. The SOD-like activity, noted $A_{SOD}$ is defined as the percentage of dismutated superoxide radicals after the 20 min.

### IV.9 - CAT mimetic activity assay

The catalytic activity of nanoceria in the disproportionation of $H_2O_2$ was assessed by spectrofluorimetry using the Amplex-Red reagent assay (Cat # A22180).[22,24,47] Briefly, 25 μL of a nanoceria dispersions in Tris-Cl buffer pH 7.5 was added to a well of a 96-well plate and mixed with 25 μL of a $H_2O_2$ solution. Then, 50 μL Amplex Red reagent/HRP working solution was added and reactions are pre-incubated for 5 minutes. Amplex Red (10-acetyl-3,7-dihydroxyphenoxazine) reaction with $H_2O_2$ catalyzed by horseradish peroxide (HRP) produces the fluorescent molecule resorufin (excitation at 571 nm and emission at 585 nm). The fluorescence was measured after incubating for 30 min with protection from light. The $H_2O_2$ concentration in each well was 5 μM, whereas those of nanoceria ranged from $3.5 \times 10^{-1}$ to $3.5 \times 10^4$ μM. The catalytic activity of nanoceria noted $A_{CAT}$ is defined as the percentage of decomposed $H_2O_2$ determined at the end of the assay.

# Acknowledgment

We thank Jérôme Fresnais, Evdokia Oikonomou, Fanny Mousseau, Chloé Puisney for fruitful discussions. Dr. Sophie Novak from the "Plateforme Rayons X" at the University Paris-Diderot is acknowledged for carrying out the wide-angle X-ray scattering technique on nanoceria. Dr. Rémi LeBorgne from the "Plateforme de microscopie électronique" at the Institut Jacques Monod is acknowledged for his assistance and support with transmission electronic microscopy. Dr. Christophe Methivier and Dr. Christophe Calers from the Laboratoire de Réactivité de Surface at the University Pierre et Marie Curie, Paris 6 are thanked for carrying out the X-ray photoelectron spectrometry (XPS) experiments. ANR (Agence Nationale de la Recherche) and CGI (Commissariat à l'Investissement d'Avenir) are acknowledged for their financial support of this work through Labex SEAM (Science and Engineering for Advanced Materials and devices) ANR 11 LABX 086, ANR 11 IDEX 05 02. This research was supported in part by the Agence Nationale de la Recherche under the contract ANR-13-BS08-0015 (PANORAMA), ANR-12-CHEX-0011 (PULMONANO) and ANR-15-CE18-0024-01 (ICONS, Innovative polymer coated cerium oxide for stroke treatment).

# References

1. I. Celardo, J. Z. Pedersen, E. Traversa and L. Ghibelli, *Nanoscale*, 2011, **3**, 1411-1420.
2. J. Gagnon and K. M. Fromm, *Eur. J. Inorg. Chem.*, 2015, **SI**, 4510-4517.
3. A. Trovarelli and J. Llorca, *ACS Catal.*, 2017, **7**, 4716-4735.
4. A. Trovarelli and P. Fornasiero, *Catalysis by Ceria and Related Materials, 2nd Edition*, Imperial College Press, London, 2013.
5. M. E. Davis and R. J. Davis, *Fundamentals of chemical reaction engineering*, McGraw-Hill Higher Education, New York, 2003.
6. C. Xu and X. Qu, *NPG Asia Materials*, 2014, **6**, e90.
7. E. G. Heckert, A. S. Karakoti, S. Seal and W. T. Self, *Biomaterials*, 2008, **29**, 2705-2709.
8. T. Pirmohamed, J. M. Dowding, S. Singh, B. Wasserman, E. Heckert, A. S. Karakoti, J. E. S. King, S. Seal and W. T. Self, *Chem. Commun.*, 2010, **46**, 2736-2736.
9. Y. Xue, Q. Luan, D. Yang, X. Yao and K. Zhou, *J. Phys. Chem. C*, 2011, **115**, 4433-4438.
10. J. M. Dowding, S. Seal and W. T. Self, *Drug Deliv. Transl. Res.*, 2013, **3**, 375-379.
11. N. Ould-Moussa, M. Safi, M.-A. Guedeau-Boudeville, D. Montero, H. Conjeaud and J.-F. Berret, *Nanotoxicology*, 2014, **8**, 799-811.

12. T. Xia, M. Kovochich, M. Liong, L. Madler, B. Gilbert, H. B. Shi, J. I. Yeh, J. I. Zink and A. E. Nel, *ACS Nano*, 2008, **2**, 2121-2134.
13. J. Y. Ma, R. R. Mercer, M. Barger, D. Schwegler-Berry, J. Scabilloni, J. K. Ma and V. Castranova, *Toxicol. Appl. Pharmacol.*, 2012, **262**, 255-264.
14. A. Thill, O. Zeyons, O. Spalla, F. Chauvat, J. Rose, M. Auffan and A. M. Flank, *Environ. Sci. Technol.*, 2006, **40**, 6151-6156.
15. M. Auffan, J. Rose, M. R. Wiesner and J. Y. Bottero, *Environ. Pollut.*, 2009, **157**, 1127-1133.
16. M. Horie, K. Nishio, H. Kato, K. Fujita, S. Endoh, A. Nakamura, A. Miyauchi, S. Kinugasa, K. Yamamoto, E. Niki, Y. Yoshida, Y. Hagihara and H. Iwahashi, *J. Biochem.*, 2011, **150**, 461-471.
17. S. Chigurupati, M. R. Mughal, E. Okun, S. Das, A. Kumar, M. McCaffery, S. Seal and M. P. Mattson, *Biomaterials*, 2013, **34**, 2194-2201.
18. F. R. Cassee, E. C. van Balen, C. Singh, D. Green, H. Muijser, J. Weinstein and K. Dreher, *Crit. Rev. Toxicol.*, 2011, **41**, 213-229.
19. A. S. Karakoti, P. Munusamy, K. Hostetler, V. Kodali, S. Kuchibhatla, G. Orr, J. G. Pounds, J. G. Teeguarden, B. D. Thrall and D. R. Baer, *Surf. Interface Analysis*, 2012, **44**, 882-889.
20. J. Chen, S. Patil, S. Seal and J. F. McGinnis, *Nat. Nanotechnol.*, 2006, **1**, 142-150.
21. J. Niu, A. Azfer, L. M. Rogers, X. Wang and P. E. Kolattukudy, *Cardiovasc. Res.*, 2007, **73**, 549-559.
22. M. Soh, D.-w. Kang, H.-g. Jeong, D. Kim, D. Y. Kim, W. Yang, C. Song, S. Baik, I.-y. Choi, S.-k. Ki, H. J. Kwon, T. Kim, C. K. Kim, S.-h. Lee and T. Hyeon, *Angew. Chemie Int. Ed.*, 2017, 1-6.
23. K. L. Heckman, W. DeCoteau, A. Estevez, K. J. Reed, W. Costanzo, D. Sanford, J. C. Leiter, J. Clauss, K. Knapp, C. Gomez, P. Mullen, E. Rathbun, K. Prime, J. Marini, J. Patchefsky, A. S. Patchefsky, R. K. Hailstone and J. S. Erlichman, *ACS Nano*, 2013, **7**, 10582-10596.
24. C. K. Kim, T. Kim, I.-Y. Choi, M. Soh, D. Kim, Y.-J. Kim, H. Jang, H.-S. Yang, J. Y. Kim, H.-K. Park, S. P. Park, S. Park, T. Yu, B.-W. Yoon, S.-H. Lee and T. Hyeon, *Angew. Chemie Int. Ed.*, 2012, **51**, 11039-11043.
25. S. Baer and G. Stein, *J. Chem. Soc.*, 1953, 3176-3176.
26. P. Sigler and B. J. Masters, *J. Am. Chem. Soc.*, 1957, **1314**, 6353-6357.
27. B. H. J. Bielski and E. Saito, *J. Phys. Chem.*, 1962, **66**, 2266-2268.
28. E. Saito and B. H. J. Bielski, *J. Am. Chem. Soc.*, 1961, **83**, 4467-4468.
29. N. Kitajima, S. Fukuzumi and Y. Ono, *J. Phys. Chem.*, 1978, **82**, 1505-1509.
30. C. Walkey, S. Das, S. Seal, J. Erlichman, K. Heckman, L. Ghibelli, E. Traversa, J. F. McGinnis and W. T. Self, *Environ. Sci. Nano*, 2015, **2**, 33-53.
31. J.-D. Cafun, K. O. Kvashnina, E. Casals, V. F. Puntes and P. Glatzel, *ACS Nano*, 2013, **7**, 10726-10732.
32. S. S. Lee, H. Zhu, E. Q. Contreras, A. Prakash, H. L. Puppala and V. L. Colvin, *Chem. Mater.*, 2012, **24**, 424-432.
33. M. Ornatska, E. Sharpe, D. Andreescu and S. Andreescu, *Anal. Chem.*, 2011, **83**, 4273-4280.
34. Y. Yang, Z. Mao, W. Huang, L. Liu, J. Li, J. Li and Q. Wu, *Sci. Rep.*, 2016, **6**, 35344.
35. D. Damatov and J. M. Mayer, *Chem. Commun.*, 2016, **52**, 10281-10284.
36. F. H. Scholes, C. Soste, A. E. Hughes, S. G. Hardin and P. R. Curtis, *Appl. Surf. Sci.*, 2006, **253**, 1770-1780.
37. Y.-J. Wang, H. Dong, G.-M. Lyu, H.-Y. Zhang, J. Ke, L.-Q. Kang, J.-L. Teng, L.-D. Sun, R. Si, J. Zhang, Y.-J. Liu, Y.-W. Zhang, Y.-H. Huang and C.-H. Yan, *Nanoscale*, 2015, **7**, 13981-13990.
38. C. Zang, X. Zhang, S. Hu and F. Chen, *Appl. Catal. B Environ.*, 2017, **216**, 106-113.
39. L. Qi, J. P. Chapel, J. C. Castaing, J. Fresnais and J.-F. Berret, *Soft Matter*, 2008, **4**, 577-585.
40. A. Sehgal, Y. Lalatonne, J.-F. Berret and M. Morvan, *Langmuir*, 2005, **21**, 9359-9364.
41. L. Qi, A. Sehgal, J.-C. Castaing, J.-P. Chapel, J. Fresnais, J.-F. Berret and F. Cousin, *ACS Nano*, 2008, **2**, 879-888.

42. S. S. Lee, W. Song, M. Cho, H. L. Puppala, P. Nguyen, H. Zhu, L. Segatori and V. L. Colvin, *ACS Nano*, 2013, **7**, 9693-9703.
43. S. Tsunekawa, K. Ishikawa, Z. Q. Li, Y. Kawazoe and A. Kasuya, *Phys. Rev. Lett.*, 2000, **85**, 3440-3443.
44. R. Klein and B. D'aguanno, in *Light Scattering, Principles and Developments*, ed. W. Brown, Oxford, 1996, pp. 30 - 102.
45. M. Das, S. Patil, N. Bhargava, J. F. Kang, L. M. Riedel, S. Seal and J. J. Hickman, *Biomaterials*, 2007, **28**, 1918-1925.
46. F. Chen, X. Shen, Y. Wang and J. Zhang, *Appl. Catal. B Environ.*, 2012, **121-122**, 223-229.
47. D. Gil, J. Rodriguez, B. Ward, A. Vertegel, V. Ivanov and V. Reukov, *Bioeng.*, 2017, **4**, 18-18.
48. Z. Tian, X. Li, Y. Ma, T. Chen, D. Xu, B. Wang, Y. Qu and Y. Gao, *ACS Appl. Mater. Interfaces*, 2017, **9**, 23342-23352.
49. A. Karakoti, S. Singh, J. M. Dowding, S. Seal and W. T. Self, *Chem. Soc. Rev.*, 2010, **39**, 4422-4422.
50. K. Y. Foo and B. H. Hameed, *Chem. Eng. Sci.*, 2010, **156**, 2-10.
51. S. G. Evans, J. M. Elliott, L. M. Andrews, P. N. Bartlett, P. J. Doyle and G. Denuault, *Anal. Chem.*, 2002, **74**, 1322-1326.
52. Y. Li, C. Sella, F. Lemaitre, M. G. Collignon, L. Thouin and C. Amatore, *Electroanalysis*, 2013, **25**, 895-902.
53. B. Liu, Z. Sun, P.-J. J. Huang and J. Liu, *J. Am. Chem. Soc.*, 2015, **137**, 1290-1295.
54. W. Huang and Y. Gao, *Catal. Sci. Technol.*, 2014, **4**, 3772-3784.
55. D. R. Mullins, *Surf. Sci. Rep.*, 2015, **70**, 42-85.
56. K. Reed, A. Cormack, A. Kulkarni, M. Mayton, D. Sayle, F. Klaessig and B. Stadler, *Environ. Sci. Nano*, 2014, **1**, 390-405.
57. B. Chanteau, J. Fresnais and J.-F. Berret, *Langmuir*, 2009, **25**, 9064-9070.
58. M. Safi, H. Sarrouj, O. Sandre, N. Mignet and J.-F. Berret, *Nanotechnology*, 2010, **21**, 145103.
59. M. Nabavi, O. Spalla and B. Cabane, *J. Colloid Interface Sci.*, 1993, **160**, 459-471.
60. C. Singh, S. Friedrichs, G. Ceccone, N. Gibson, K. A. Jensen, M. Levin, H. G. Infante, D. Carlander and K. Rasmussen, *Cerium Dioxide, NM-211, NM-212, NM-213. Characterisation and test item preparation*, European Commission Joint Research Centre Institute for Health and Consumer Protection, Ispra, Italy, 2014.
61. E. K. Oikonomou, F. Mousseau, N. Christov, G. Cristobal, A. Vacher, M. Airiau, C. Bourgaux, L. Heux and J. F. Berret, *J. Phys. Chem. B*, 2017, **121**, 2299-2307.
62. J.-F. Berret, O. Sandre and A. Mauger, *Langmuir*, 2007, **23**, 2993-2999.

TOC image

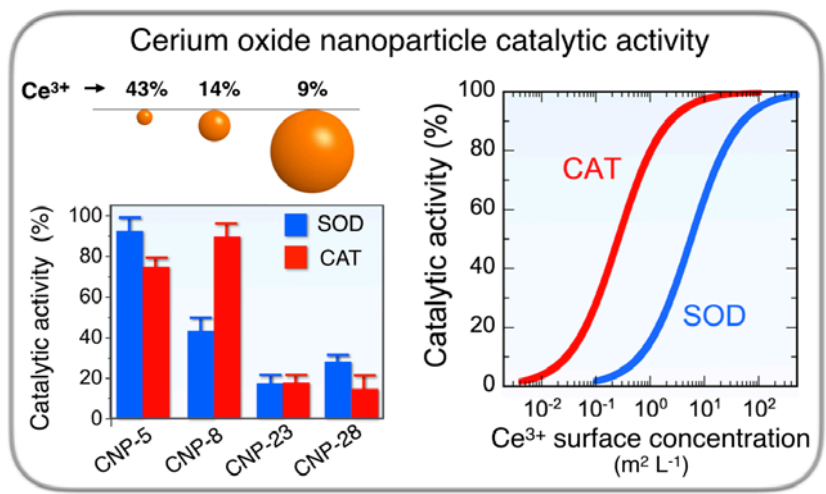